\documentclass[a4paper]{article}
\usepackage{a4wide}
\usepackage[centertags]{amstex}
\usepackage{amssymb}
\usepackage{epsfig}             

\makeatletter
\renewcommand\section{\@startsection{section}{1}{\z@}%
                                   {-3.5ex \@plus -1ex \@minus -.2ex}%
                                   {2.3ex \@plus.2ex}%
                                   {\reset@font\Large\scshape}}
\renewcommand\subsection{\@startsection{subsection}{2}{\z@}%
                                     {-3.25ex\@plus -1ex \@minus -.2ex}%
                                     {1.5ex \@plus .2ex}%
                                     {\reset@font\large\slshape}}

\makeatother

\def\ap#1#2#3   {{Ann.\ Phys.\ (NY)}\ #1 (#2) #3}
\def\ib#1#2#3   {ibid., #1 (#2) #3}
\def\np#1#2#3   {{Nucl.\ Phys.}\ #1 (#2) #3}
\def\npbps#1#2#3   {{Nucl.\ Phys.\ B (Proc.\ Suppl.)}\ #1 (#2) #3}
\def\pl#1#2#3   {{Phys.\ Lett.}\ #1 (#2) #3}
\def\prep#1#2#3 {{Phys.\ Rep.}\ #1 (#2) #3}
\def\prev#1#2#3 {{Phys.\ Rev.}\ #1 (#2) #3}
\def\prl#1#2#3  {{Phys.\ Rev.\ Lett.}\ #1 (#2) #3}
\def\zp#1#2#3  {{Z.\ Phys.}\ #1 (#2) #3}

\newcommand\half{\ensuremath{\frac{1}{2}}}
\newcommand\Tr{\operatorname{Tr}}
\newcommand\GL{\ensuremath{\mathcal{G}}}
\newcommand\diag{\operatorname{diag}}

\newcommand\OpW{\ensuremath{\mathcal{W}}}
\newcommand\OpM{\ensuremath{\mathcal{M}}}
\newcommand\OpE{\ensuremath{\mathcal{E}}}

\newlength{\figwidth}
\newlength{\figwidthless}
\newcounter{hours}\newcounter{minutes}

\begin{document}

\thispagestyle{empty}

\begin{center}
January 1999 \hfill \large IFUP-TH 5/99 \\
\noindent \hfill \large hep-lat/9902002
\end{center}

\vspace{1cm plus 1fil}

\begin{center}

\textbf{\huge Breaking of the adjoint string in 2+1 dimensions}

\vspace{1cm plus 1fil}

{\large
{\bfseries P.W. Stephenson \\}
           Dipartimento di Fisica dell'Universit\`a and INFN, \\
           I-56100 Pisa, Italy \\
           Email: \texttt{pws@@ibmth.df.unipi.it}}

\vspace{1cm plus 1fil}

\textbf{\Large Abstract}

\end{center}

\noindent The roughly linear rise of the potential found between
adjoint sources in SU($N$) in lattice simulations is expected to
saturate into a state of two `gluelumps' due to gluonic screening.
We examine this in SU(2) in 2+1 dimensions.  Crossover between
string-like and broken states is clearly seen by the mixing-matrix
technique, using different operators to probe the two states; the
breaking behaviour is rather abrupt.  Furthermore, we are able to show
that both types of operator have a finite overlap with both states;
in the case of the Wilson loops the overlap with the broken string
is, as predicted, very small.

\vspace{1cm}



\newpage

\section{Introduction}

The potential $V(R)$ between sources a distance $R$ apart in SU($N$)
gauge theory has been accurately calculated on the lattice for various
$N$ and in two and three dimensions and forms a vital ingredient in
our understanding of quark confinement.  In the case of the potential
between fundamental sources, i.e.\ a quark and an antiquark, one
expects the potential to saturate due to pair creation in the vacuum;
in the case of sources in the adjoint representation of SU($N$), the
potential should also saturate, this time due to screening by gluonic
fields.  Seeing both these effects is an important step to
understanding confinement phenomenology.

However, until recently both have proved elusive.  To see any kind of
screening of the fundamental potential requires a full QCD simulation
with sea quarks, which is still difficult.  The best simulations of
real-world QCD to date~\cite{CP98} still have fairly heavy quarks and
standard potential calculations do not show a clear sign of string
breaking.

Two ways have been suggested of improving the situation.  Rather than
simply asking if the adjoint string breaks, one needs to know: (i)
whether states for both broken and unbroken strings exist; (ii) what
is the ground state of $V(R)$ as $R$ increases; (iii) what happens if,
as we expect, the states cross over.  So one should ensure one has a
suitable basis of operators for both the broken and unbroken
string~\cite{Gu98}.  Since Wilson loops, the standard probe for the
gluonic field, appear to probe the broken string only very poorly, and
conversely operators more resembling the broken string are not good
probes for the complete one, one can look at mixing between these
operators to try and reconstruct the real ground state, using the
diagonalisation method originally proposed in ref.~\cite{GrMi83}.  In
the case of string breaking, success has been obtained in various
models \cite{PhWi98,KnSo98}, although not yet in full
3+1-dimensional QCD.

Secondly, one can look at Wilson loop operators in a regime where they
may be better able to resolve the broken state.  By the use of
improved actions, enabling the use of coarser lattices, together with
a space--time asymmetry, enabling a short time separation for
resolving effective masses, Trottier has for the first time found
unambiguous evidence of string breaking in full 2+1-dimensional QCD
without resorting to mixing matrices~\cite{Tr98}.

Our intention here is to apply these ideas to the case of the adjoint
potential.  In this case string breaking was seen using the
mixing-matrix approach some time ago in 3+1 dimensions by
Michael~\cite{Mi92}, but a more detailed examination is certainly due.
In particular, little evidence was then found for mixing between the
broken and unbroken states.  In this work we perform simulations in 2+1
dimensions, where the statistics are more in our favour but the physics
of confinement appears similar to the 3+1-dimensional case~\cite{Te98}.
A study using Wilson loop operators without taking account of mixing
found essentially no sign of the broken string~\cite{PoTr95}.  Here we
shall use a more complete basis of operators and suggest that when we do
this there is little mystery about the breaking of the adjoint string.

In the next section, we describe our method; then we present some
initial results on the scaling behaviour of the string tension and of
gluelumps to help us choose parameters for the main simulations.  The
main results are presented in the fourth section; finally we give a
brief summary of our conclusions.

\section{Method}

We consider standard SU(2) lattice gauge theory, but in addition to
the Wilson action we have also used the asymmetric improved
action~\cite{Mo96}:
\begin{equation}
  S \equiv \beta \xi \big( \frac{5}{3u_s^4}W_\mathrm{sp} +
  \frac{4}{3\xi^2u_s^2u_t^2}W_\mathrm{tp} - \frac{1}{12u_s^6}W_\mathrm{sr}
  - \frac{1}{12\xi^2u_s^4u_t^2}W_\mathrm{str}\big).
  \label{eq:spimp}
\end{equation}
$W$ are various forms of Wilson loop, $W_c = 1/2 \Tr U_c$ where $U_c$
is the product of link matrices corresponding to the loop shape $c$. The
shapes are, respectively, spatial
plaquettes (all links in one of the two space directions), temporal
plaquettes (one pair of links in the time direction), spatial
rectangles, and short temporal rectangles where the single lattice
spacing link is always in the time direction.  $\xi$ is the
anisotropy, i.e. the ratio of lattice spacings $a_t/a_s$, and $u_s$
and $u_t$ are factors inserted for a mean field improvement; as
justified below, we have in fact set both of these to unity in our
main simulations.  As usual, $\beta$ is the inverse coupling $4/g^2$.
This action has errors of $\mathcal{O}(a_t^2)$, which are small
because they are suppressed by a factor $\xi^4$.  The tree level
action $u_s = u_t = 1$ has errors $\mathcal{O}(a_s^2g^2)$; the
mean field improvement is designed to reduce as much as possible terms
of the form $\mathcal{O}(a_s^2g^n)$.  The standard Wilson action is
recovered by taking just the first two terms in the parentheses with
coefficient $\beta$.  From now on, we take $a\equiv a_s$ as the basic
lattice spacing.

A simpler form of improvement is to take the temporal part to have the
same coefficients as the spatial part, i.e.\ $5/3$ for all plaquettes
and $1/12$ for all rectangles, as done in ref.~\cite{Tr98}.  We shall
refer to this as ST-improvement (for both space and time) to
distinguish it from the S-improvement of eqn.~(\ref{eq:spimp}).  The
latter has the advantage that only single links occur in the time
direction, which keeps the transfer matrix simple.  The effects of the
higher modes in the ST-form are most apparent when looking at the
overlaps of loop operators with physical states, as the
coefficients are no longer positive definite.  In this work the
overlaps are particularly important, so all our improved results
use the S-type.

Updating of the lattice was done by a Kennedy--Pendleton
heatbath~\cite{KePe85} with (after equilibration) four sweeps of exact
over-relaxation between each heatbath sweep.

\begin{figure}
  \begin{center}
    \psfig{file=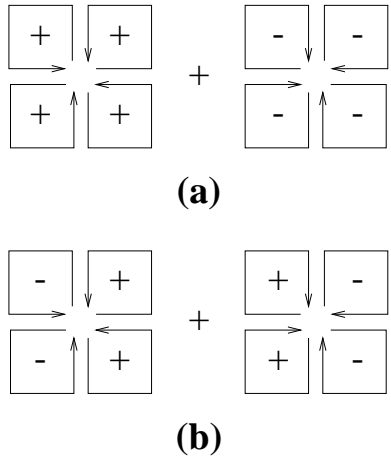,angle=270,width=2in}
  \end{center}
  \caption{The operators used for probing (a) a magnetic and (b)
    an electric gluelump.  In each case, the fields start and finish at
    the same site and all the squares are open plaquettes.}
  \label{fig:clover}
\end{figure}

To look at both broken and unbroken strings, we use different sets of
operators.  All quantities in the following discussion are in lattice
units.  For our broken string operators, the method follows that
originally described in full in 3+1 dimensions in ref.~\cite{Mi85} and
used in early work on the adjoint sector~\cite{MiXX}.  As briefly
mentioned above, the broken adjoint string is expected to resemble a
pair of gluelumps, each of which is a localised static gluonic field
surrounded by dynamical gluons.  This is because of the static spatial
part of the Wilson loop; in the real world, no such source exists and
the pure glue states are glueballs, i.e.\ states with an indeterminate
number of purely dynamical gluons.  However, gluelumps, when they can
form, are considerably lighter.  In two spatial dimensions, there are
two types of gluelump with distinct quantum numbers, `magnetic' and
`electric', whose masses were calculated in ref.~\cite{PoTr95} by
essentially the method we are adopting.  Suitable operators can be
made from combinations $G_i$ ($i$ standing for magnetic or electric)
of `open' plaquettes (i.e.\ the SU(2) matrix corresponding to the
complete path before the trace is taken) as shown in
fig.~\ref{fig:clover}.  We could in principle use larger Wilson loops
than plaquettes, but results are entirely satisfactory when combining
the use of plaquette operators with the fuzzing prescription described
later.

The operators $G_i$ are traceless, as the individual parts have $C=-1$
and the trace of an SU(2) matrix is real so has only $C=+1$.  Thus
before we can measure anything we need to combine the two parts in a
gauge-invariant way using temporal links.  Finally, we need a trace in
the adjoint representation.  More formally, the complete operator
\GL~\cite{Mi85} consists of
\begin{equation}
  \GL_i(T) = \Tr(G_i(0)\sigma^a)A^{ab}(U(0,T))\Tr(G_i^\dag(T)\sigma^b),
\end{equation}
where $U(0,T)$ is the product of links between $G_i(t=0)$ and
$G_i(t=T)$; $t$ and $T$ will be taken in lattice units, i.e.\ they count
temporal lattice spacings.   For the adjoint projection $U$ is combined
in the form,
\begin{equation}
  A^{ab}(U) = \half\Tr(\sigma^aU\sigma^bU^\dag).
\end{equation}
Here, the $\sigma^a$ are the Pauli matrices, $a=1\dots3$, and the
standard identity
\begin{equation}
  \sigma_{ij}^a\sigma_{kl}^b = 2(\delta_{il}\delta_{jk} -
  \half\delta_{ij}\delta_{kl})
\end{equation}
means that \GL\ has the simple form
\begin{equation}
  \GL_i(T) = 2\Tr(G_i(0)U(0,T)G_i^\dag(T)U^\dag(0,T)),
  \label{eq:GL}
\end{equation}
since as already indicated the traces of the $G_i$ vanish.
This `dumbbell' operator, corresponding to creation of a gluelump and
its annihilation after time $T$, allows us to measure the gluelump
mass in the usual way.  The broken string state is
modelled by two of these $\GL_i$ separated by some distance $R$;
for large enough $R$ in the confined phase this will overlap
predominantly onto a state of twice the mass of the individual gluelump.
We shall refer to it in what follows as the `two-dumbbell operator',
to distinguish from the physical state of two gluelumps that we expect
it to resemble at least for large $R$.  We shall also use the
shorthand `two magnetic dumbbells' which is to be interpreted as two
such operators, each with the quantum numbers of a magnetic gluelump.

The operator for the adjoint potential from Wilson loops
$W_\mathrm{adj}(R,T)$ corresponding to the unbroken string distance
$R$ apart propagating for time $T$ comes simply from calculating the
loop in the usual way then taking the adjoint trace,
\begin{align}
  \Tr_\mathrm{adj} U_c &\equiv \frac{1}{3}(4\Tr_\mathrm{fund}^2 U_c - 1), \\
  \Tr_\mathrm{fund} U_c &\equiv \half \Tr U_c,
  \label{eq:traces}
\end{align}
where $U_c$ is the matrix product of the links corresponding to the
closed path $c$.  It is thus simple to keep the fundamental
potential for comparison.

\begin{figure}
  \begin{center}
    \psfig{file=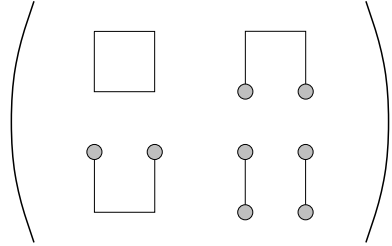,angle=270,width=2in}
  \end{center}
  \caption{Schematic depiction of the states in the mixing matrix
    $M(R,T)$; each filled circle represents one of the operators of
    fig.~\protect\ref{fig:clover}, and all links are taken in the
    adjoint representation.  In each state shown $R$ is horizontal and
    $T$ vertical.}
  \label{fig:mixing}
\end{figure}

Finally, to calculate mixing between the states we follow exactly the
method of ref.~\cite{PhWi98}, but replacing the scalar meson operators
with the operators $G_i$ and the fundamental with adjoint fields.  Then
we need operators corresponding to an unbroken string developing into a
two-gluelump state or vice versa.  These can be seen in the off-diagonal
parts of the matrix in fig.~\ref{fig:mixing}.  They are identical to the
operators $G_i$, except with a different shape to the links, so we again
use eqn.~(\ref{eq:GL}) with $U(0,T)$ replaced by the appropriate
`U'-shaped product of links, and with the $G_i$ spatially instead of
temporally separated.

All our gluonic operators for measurement use the now standard
iterative `fuzzing' prescription~\cite{APE87}, in which a spatial link
is replaced recursively by a sum of the original link and the four
three-link spatial staples around it, preserving the lattice symmetry.
Our conventions for the constants are, in condensed form,
\begin{equation}
  U^{(n_f+1)}(x,\mu) = P\lbrace c_f U^{(n_f)}(x,\mu) + \sum_{\nu\ne\mu}
  \mathrm{staples}^{(n_f)}(x,\mu,\pm\nu)\rbrace
\end{equation}
where $P$ represents a projection back to SU(2) for numerical
convenience, the staples are the four matrices representing the
possible three-link paths from position $x$ to $x+\hat\mu$ going first
in either the $+\nu$ or $-\nu$ direction, and $U^{(n_f)}(x,\mu)$ is a
link matrix after $n_f$ iterations of fuzzing,
where the original links are those for $n_f=0$.
The choice of the coefficient of the original link $c_f$ 
and the number of iterations need to be tuned to maximise
the overlap to physical states.  In addition to improving the
overlap, this also provides us with different trial operators for the
mixing matrix.

The resulting mixing matrix $M(R,T)$ is shown schematically in
fig.~\ref{fig:mixing}; different fuzzing levels are also included for
each type of operator.  The off-diagonal terms in the full matrix
consist in taking the appropriate product of links or pair of local
operators at $t=0$ and at $t=T$: for example, the pure Wilson loop part
of the matrix includes an unfuzzed product of links at $t=0$ together
with a fuzzed product at $t=T$, and so on.  Note that we always use a
given fuzzing level for a given timeslice, so that both gluelumps in the
off-diagonal operators of fig.~\ref{fig:mixing} always have the same
fuzzing.

We then need to diagonalise the corresponding transfer
matrix~\cite{GrMi83}. Ideally, we wish to have
\begin{multline}
  M'(R,T+1) \lvert\psi(R)\rangle = \diag(C_0(R)\exp(-m_0(R)a_t),\quad
  C_1(R)\exp(-m_1(R)a_t),\quad\dots) \\
  M'(R,T) \lvert\psi(R)\rangle
  \label{eq:diag}
\end{multline}
for the spectrum of states $\psi_i(R)$ with masses $m_i(R)$ and some
coefficient $C_i(R)$.  Actually, we have only a truncated basis and
finite statistics and can only hope for an approximation to this, which
we arrange by solving the corresponding eigenvalue problem for the
original $M(R,T)$ and $M(R,T+1)$ for each $R$ separately.  Thus we need
to pick a basis $(T_D,T_D+1)$ for the process.  Various factors come
into this: first, operators have smaller relative errors at lower $T_D$;
second, the contaminating effect of higher states which cannot fit into
our finite basis is less at larger $T_D$; finally there may be low-lying
states with small coefficients $C_i(R)$ which are swamped at small
$T_D$, but which are clearer when we go to larger $T_D$.

We use the notation $C(R,t)$ to indicate the set of diagonalised
correlation functions,
\begin{equation}
  C_i(R,t) = C_{i,0}(R) \exp(m_0(R,t)) + C_{i,1}(R)
  \exp(m_1(R,t)) + \dots
\end{equation}
where the subscript $i$ is an index for the eigenvalue taken in the
diagonalisation.  After the diagonalisation, only one $C_{i,j}(R)$
should dominate for each, at least for the lowest states, and we order
the eigenvalues so that this is $C_{i,i}(R)$.  We consider the first
two eigenmodes in each case, which (barring problems in the numerical
procedure) give the lowest 
two masses $m_0(R)$ and $m_1(R)$.  We then fit a single exponential to
$C_i(R,t)$ starting from some $t$ to be decided later: all our fits
use a correlated $\chi^2$, where errors are determined by the
bootstrap method.  If the fit is acceptable, then the calculated
coefficient $C_{i,i}(R)/C_i(R,0)$ gives an estimate of the overlap
of the operator basis with the physical state, something
of particular interest in the present case; this is the quantity meant
when we talk of overlaps in what follows.  From basic quantum
mechanics the overlaps onto the basis of states are positive
and sum to one (again, barring numerical problems) for both our
actions. We also define the effective mass for a given $R$ in the
usual way,
\begin{equation}
  m_\mathrm{eff}(t) = - \log(C(t)/C(t+1)),
\end{equation}
a useful indication of whether a single exponential term dominates.
However, it should be noted that the masses quoted are from the fit to
the exponential, not effective masses, to allow us a reasonable
estimate of the overlap.

For the temporal links in the Wilson (but not the improved) case we
have used the so-called `multihit' technique~\cite{PaPe83} to improve
statistics, though in the case of SU(2), where the improved link can
be calculated exactly, the name `single bullseye' would be more
appropriate.  In this method we can perform an analytic integration on
half of the temporal links (all those which have no plaquette in
common), replacing them by
\begin{equation}
  U_l \longrightarrow U_l' = \frac{I_2(\beta k_l)}{I_1(\beta k_l)} V_l,
  \label{eq:bullseye}
\end{equation}
where $I_n(x)$ is a modified Bessel function, $k_l V_l$ is the sum of the
four three-link `staples' surrounding the original link $U_l$ with the
same start and end point such that $k_l$ is a positive real number and
$V_l$ is a (uniquely determined) SU(2) matrix.  Expressions involving
the right hand side of eqn.~(\ref{eq:bullseye}) have the same
expectation value but with less noise.  Note that this expression is
for the fundamental representation; for the adjoint case one can use
the same improved matrices but when taking an adjoint trace as in
eqn.~(\ref{eq:traces}) instead of subtracting $1/3$ from
$4\Tr_\mathrm{fund}^2(U_c)/3$, one must subtract $\Tr({U_c'}^\dag
U_c')/6$; this is easily implemented.  There is also a particularly
simple method for calculating the $I_n(x)$ required, namely via the
recurrence relation
\begin{equation}
I_{n-1}(x) = I_{n+1}(x) + \frac{2n}{x}I_n(x),
\end{equation}
using the fact that the $I_n(x)$ decrease with increasing $n$.  Starting
from some high $n$ (we have found $n=17$ sufficient for 32-bit
precision), we take $I_{n+2}(x) = 0$ and $I_{n+1}=\epsilon$ for some
arbitrary but small $\epsilon$, and work backwards.  This gives $I_2(x)$
and finally $I_1(x)$ up the same overall constant factor, so we simply
divide the results obtained for the two to obtain $I_2(x)/I_1(x)$.  In
the S-improved case, where the action is more complicated and more links
are involved, we have not attempted to modify the multihit algorithm
appropriately and so have not used it.

\section{Choice of actions and lattices}

We wish to see if coarse lattices are as useful as they are in the
case of breaking of the fundamental string~\cite{Tr98}.  For this we
would first like some idea of the behaviour of the S-improved actions
in this region.  It is customary to use the fundamental string tension
$K$ for setting the scale, which has a proper continuum limit.  In the
Wilson case there is already ample data for this~\cite{Te98}.  We also
calculate the masses of the two lowest-lying gluelumps with both
magnetic and electric quantum numbers for comparison.  As a static
source, these have logarithmic divergences~\cite{LaPh97}, so that
there is no continuum limit for the bare gluelump masses, nor is there
a simple scaling behaviour.  However, the adjoint potential in which
we are interested consists of two such static sources and hence twice
the gluelump mass is an aid to matching the potential at different
$\beta$.

In 2+1 dimensions, the fundamental potential has a long range
Coulombic part which is logarithmic, making it difficult to separate
out the linear effect of the string tension.  We have followed
ref.~\cite{Te98} and determined $K$ from correlations of the Polyakov
loop $\Tr U_P(t)$, which wraps around the spatial boundary at some $t$ and
hence has length $N_s$ units on an $N_s^2N_t$ lattice.  Including the
appropriate form of the universal string
correction~\cite{LuSy80,FoSc85}, the asymptotic form is
\begin{equation}
  \langle \Tr U_P(0) \Tr U_P(T) \rangle @>t\to\infty>> \exp -\left(Ka -
  \frac{\pi}{6aL}\right)\xi at,
\label{eq:poly}
\end{equation}
where $K$ is the string tension and the lattice
spacing $a$ and anisotropy $\xi$ have been shown explicitly.

For the improved action we have concentrated on small $\beta$, and for
the Wilson action on larger $\beta$, reflecting the
regions where we expect to use each action.  The effect of $u_t$ is
expected to be slight and we have set it everywhere to unity.  In
choosing lattice sizes, we have had an eye to the results of
ref.~\cite{Te98}, and in particular fig.~9 which appears to indicate
that a lattice dimension of $24$ at $\beta=9$ is sufficient to
eliminate finite size effects in the low-lying glueball spectrum.  We
have assumed approximate asymptotic scaling of the string tension,
which is obeyed closely enough for this purpose, to adjust lattice
sizes to have roughly this physical volume as a minimum.

\begin{table}
\begin{center}
\begin{tabular}{|rrrr|ll|}
  \hline
  $\beta$ & $\xi$ & Size & $u_s^4$ & $Ka^2$ & $\beta\sqrt{\!K}$ \\
  \hline
  2.0 & 0.25 & $6^2\times32$  & & 0.52(2)   & 1.44(3) \\
  2.0 & 0.25 & $8^2\times32$  & & 0.48(4)   & 1.38(5) \\
  2.0 & 0.25 & $12^2\times32$ & & 0.50(13)  & 1.4(2)  \\
  \hline
  2.5 & 0.25 & $8^2\times32$  & & 0.303(1)  & 1.38(2) \\
  2.5 & 0.25 & $10^2\times32$ & & 0.321(16) & 1.42(4) \\
  2.5 & 0.25 & $12^2\times32$ & & 0.33(2)   & 1.44(5) \\
  \hline
  3.0 & 0.25 & $8^2\times32$  & & 0.226(8)  & 1.43(3) \\
  3.0 & 0.25 & $12^2\times32$ & & 0.23(1)   & 1.43(3) \\
  3.0 & 0.25 & $16^2\times32$ & & 0.217(15) & 1.40(4) \\
  \hline
  3.0 & 0.333  &  $8^2\times32$ & & 0.214(11) & 1.39(4) \\
  3.0 & 0.333  & $12^2\times32$ & & 0.236(13) & 1.46(4) \\
  3.0 & 0.333  & $16^2\times32$ & & 0.23(2)   & 1.43(8) \\
  \hline
  2.0 & 0.25 &  $6^2\times32$ & 0.605 & 0.304(7)   & 1.103(13) \\
  2.0 & 0.25 &  $8^2\times32$ & 0.605 & 0.273(9)   & 1.044(16) \\
  2.0 & 0.25 & $12^2\times32$ & 0.605 & 0.37(17)   & 1.2(3)    \\
  \hline
  3.0 & 0.25 &  $8^2\times32$ & 0.767 & 0.160(5)   & 1.20(2)   \\
  3.0 & 0.25 & $12^2\times32$ & 0.767 & 0.167(5)   & 1.23(2)   \\
  3.0 & 0.25 & $16^2\times32$ & 0.767 & 0.170(8)   & 1.24(3)   \\
  \hline
\end{tabular}
\end{center}
\caption{String tensions in lattice and asymptotic units extracted
  using eqn.~(\protect\ref{eq:poly}) on various lattices with the
  S-improved action of eqn.~(\protect\ref{eq:spimp}).}
\label{tab:poly}
\end{table}

\begin{table}
\begin{tabular}{|rrrrr|ll|ll|}
  \hline
  $\beta$ & Action & $\xi$ & Size & $u_s^4$ & $M_\mathrm{gm}^{(0)}$ &
  $M_\mathrm{gm}^{(1)}$ & $M_\mathrm{ge}^{(0)}$ & $M_\mathrm{ge}^{(1)}$ \\
  \hline
  3.0 & W & 0.5 & $16^2\times32$ & & 1.95(2) & 2.91(3) & 2.26(3) & 3.14(14) \\
  5.0 & W & & $16^3$         & & 1.192(6) & 1.71(3)   & 1.466(13)& 1.81(4) \\
  6.0 & W & & $32^3$         & & 0.996(3) & 1.422(6)  & 1.235(3) & 1.55(3) \\
  7.0 & W & & $24^2\times32$ & & 0.851(4) & 1.21(3)   & 1.09(3) & 1.40(4) \\
  8.0 & W & & $24^2\times32$ & & 0.747(5) & 1.072(16) & 0.95(2) & 1.16(6) \\
  9.0 & W & & $24^2\times32$ & & 0.670(3) & 0.96(3)   & 0.81(2) & 1.21(15) \\
  \hline
  2.0 & SI & 0.25 & $6^2\times32$  & & 2.508(16) & 3.96(6) & 2.82(2) &
  3.97(6) \\
  2.0 & SI & 0.25 & $8^2\times32$  & & 2.504(8)  & 3.90(3) & 2.82(3) &
  4.06(3) \\
  2.0 & SI & 0.25 & $10^2\times32$ & & 2.508(12) & 3.96(3) & 2.828(2) &
  4.00(4) \\
  2.0 & SI & 0.25 & mean           & & 2.504(8) & 3.93(2)  & 2.824(8) &
  4.02(2) \\
  2.5 & SI & 0.25 & $8^2\times32$  & & 1.976(8) & 3.02(2)  & 2.364(8) &
  2.97(8) \\
  2.5 & SI & 0.25 & $10^2\times32$ & & 1.972(8) & 3.00(2)  & 2.364(8) &
  3.13(2) \\
  2.5 & SI & 0.25 & $12^2\times32$ & & 1.982(5) & 2.90(5)  & 2.372(8) &
  3.124(12) \\
  2.5 & SI & 0.25 & mean           & & 1.978(4) & 3.004(12) & 2.368(4) &
  3.124(8) \\
  3.0 & SI & 0.25 & $8^2\times32$  & & 1.660(8) & 2.48(2)   & 2.04(2) &
  2.65(2) \\
  3.0 & SI & 0.25 & $12^2\times32$ & & 1.664(8) & 2.516(8)  & 2.051(6) &
  2.664(8) \\
  3.0 & SI & 0.25 & $12^2\times36$ & & 1.656(3) & 2.476(8)  & 2.041(6) &
  2.664(8) \\
  3.0 & SI & 0.25 & $16^2\times32$ & & 1.648(4) & 2.472(8)  & 2.052(4) &
  2.640(12) \\
  3.0 & SI & 0.25 & mean           & & 1.653(2) & 2.488(4)  & 2.050(3) &
  2.652(4) \\
  3.0 & SI & 0.33333 & $8^2\times32$  & & 1.659(9) & 2.50(2)   & 2.046(12) &
  2.66(2) \\
  3.0 & SI & 0.33333 & $12^2\times32$ & & 1.656(6) & 2.490(12) & 2.046(6) &
  2.66(2) \\
  3.0 & SI & 0.33333 & $16^2\times32$ & & 1.660(4) & 2.499(9)  & 2.08(2)  &
  2.77(5) \\
  3.0 & SI & 0.33333 & mean           & & 1.659(3) & 2.496(6)  & 2.049(6) &
  0.888(4) \\
  4.0 & SI & 0.33333 & $16^2\times36$ & & 1.278(6) & 1.926(12) & 1.64(2)  &
  0.734(9) \\
  5.0 & SI & 0.5     & $16^2\times32$ & & 1.046(4) & 1.58(2)   & 1.328(6) &
  1.84(2)  \\
  \hline
  2.0 & SI & 0.25    & $6^2\times32$  & 0.605 & 2.028(12) & 3.28(12) & 
  2.47(3) & 3.1(2) \\
  2.0 & SI & 0.25    & $8^2\times32$  & 0.605 & 2.004(12) & 3.22(2)  &
  2.47(2) & 3.28(6) \\
  2.0 & SI & 0.25    & $12^2\times32$ & 0.605 & 2.032(12) & 3.16(2)  &
  2.49(2) & 3.36(4) \\
  2.0 & SI & 0.25    & mean           & 0.605 & 2.020(8)  & 3.21(2) & 2.480(12)
  & 3.316(12) \\
  3.0 & SI & 0.25    & $8^2\times32$  & 0.767 & 1.528(12) & 2.48(4) &
  1.93(2) & 2.76(3) \\
  3.0 & SI & 0.25    & $12^2\times32$ & 0.767 & 1.508(8)  & 2.50(2) &
  1.972(8) & 2.68(4) \\
  3.0 & SI & 0.25    & $16^2\times32$ & 0.767 & 1.528(8)  & 2.46(2) &
  1.964(12) & 2.72(3) \\
  3.0 & SI & 0.25    & mean           & 0.767 & 1.520(4)  & 2.48(2) &
  1.968(8) & 2.72(2) \\
  4.0 & SI & 0.33333 & $16^2\times32$ & 0.7735 & 1.176(6) & 1.79(2) &
  1.497(12) & 2.03(4) \\
  5.0 & SI & 0.5     & $16^2\times32$ & 0.824 & 0.986(2) & 1.44(2) & 1.256(6)
  & 1.68(4) \\
  \hline
\end{tabular}
\begin{center}
\end{center}
\caption{Gluelump masses for Wilson action
  (W) and for the S-improvement of eqn.~(\protect\ref{eq:spimp}) (SI)
  with and without mean field improvement ($u_s$). The two
  lowest-lying magnetic and the two lowest-lying electric gluelumps are shown.}
\label{tab:glmasses}
\end{table}

Results for the string tension are shown in table~\ref{tab:poly}.  In
most cases they are for about 1000 configurations, though more were
available from the runs to be described in the main section on adjoint
string breaking.  Where $u_s\ne1$, we adjusted it in a self-consistent
fashion until the spatial plaquette was $u_s^4$ to the accuracy shown;
if it is not shown, $u_s=1$.

All mass values are given in spatial units; to convert from the
temporal units in the correlator we have simply divided by the
tree-level asymmetry $\xi$ given in the second column.  This is not
necessarily correct, as there could be renormalisations to $\xi$.
However, in our later calculations we will only require masses
consistent on a given set of lattices, and the factor $\xi$ appears
consistenly in front of all masses in exponentials of the form
$\exp(-ma\xi t)$, so this is not a major concern for us.

If asymptotic scaling holds, then $\beta$, which in 2+1 dimensions has
the dimensions of length, also scales; we test this in the final
column of the results by constructing the dimensionless quantity
$\beta\sqrt{\!K}$.  It was found for the Wilson case with $\beta$ down
to around 6 that the results fit a single linear violation of
asymptotic scaling~\cite{Te98}, with an intercept at $1/\beta=0$ of
$1.341(7)$, while the ST-improvement of ref.~\cite{Tr98} shows
virtually no sign of any violation of asymptotic scaling even without
mean-field improvement down to $\beta=2$; this effect, much more
striking than in 3+1 dimensions, may be related to the fact that the
theory in our case is superrenormalisable.

\begin{figure}
  \begin{center}
    \psfig{file=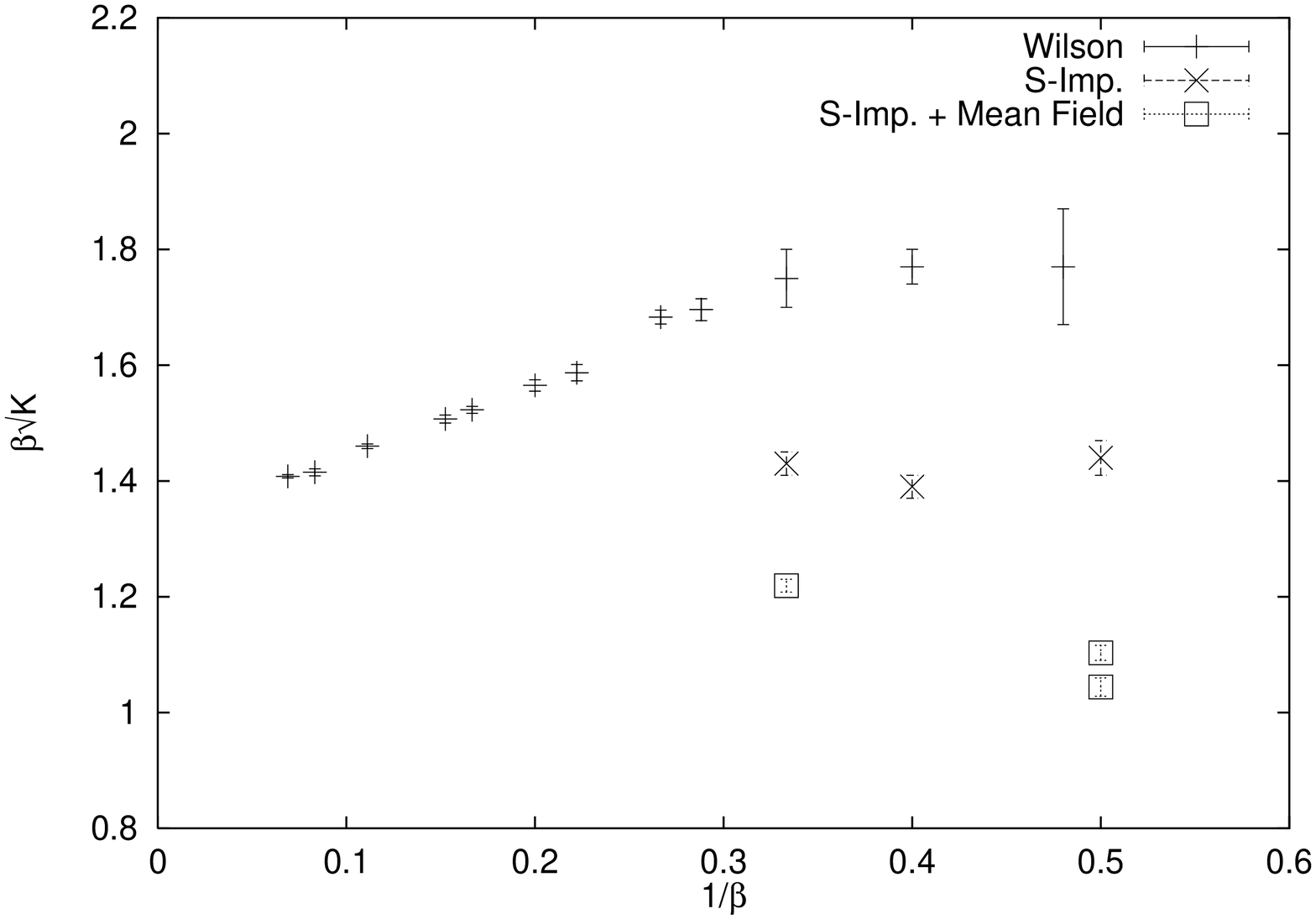,angle=270,width=\figwidthless}
  \end{center}
  \caption{The string tension $K$ for various actions, shown in
    units $\beta \sqrt{K}$ against $1/\beta$.  The Wilson
    results are from ref.~\protect\cite{Te98}.  Note that the
    tree-level asymmetry $\xi$ has been used to calculate the improved
    masses; this may undergo renormalization.}
  \label{fig:st}
\end{figure}

These results are plotted against $1/\beta$ in fig.~\ref{fig:st}.
Where the results are consistent, only a weighted mean of results from
different lattice volumes is shown.  Those without the mean-field
improvement, i.e.\ $u_s=1$, show no clear trend in either volume or
$\beta$ and cluster around $\beta\sqrt{\!K}=1.40$ to $1.43$.  This
suggests a roughly universal behaviour, which, however, is slightly
above the value $1.341(7)$ quoted above.  This may be due to the
renormalization in $\xi$ mentioned above, which is found~\cite{Mo99}
to be large in the case without mean-field improvement, so that the
real values would be shifted downwards towards the weak coupling
limit.  However, the presence of approximate scaling apart from this
shift is pleasing.

With mean-field improvement, the values are lower, and might already
be nearer the weak coupling limit.  At $\beta=2.0$ the values for the
two smaller volumes are not entirely consistent (the largest volume is
not plotted), although this could still be a statistical fluctuation.

\begin{figure}
  \begin{center}
    \psfig{file=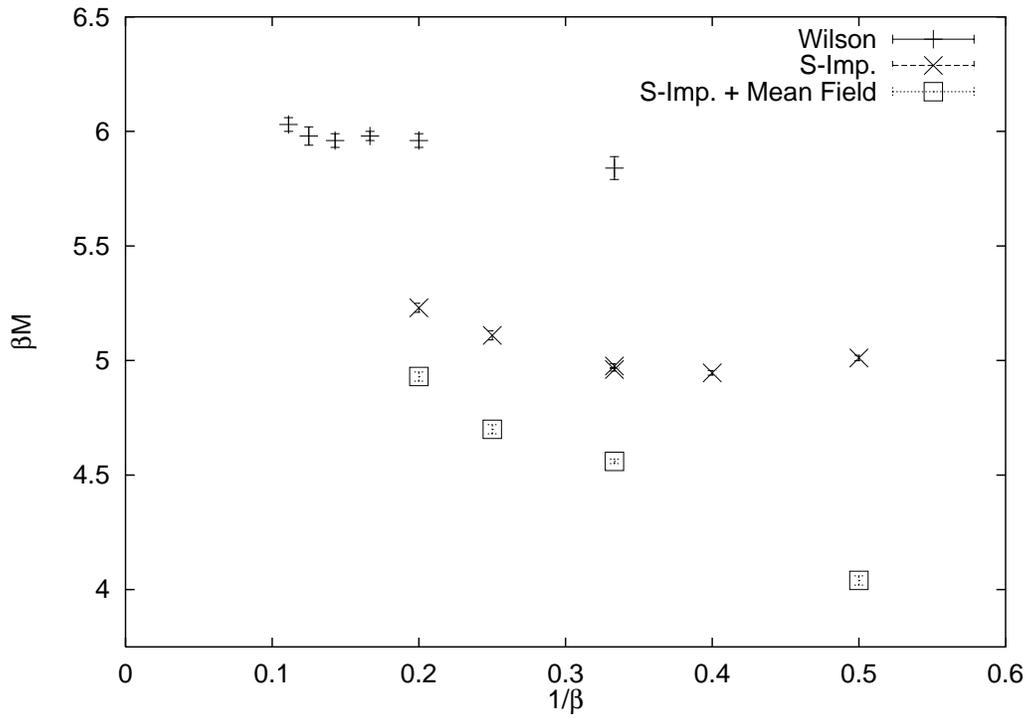,angle=270,width=\figwidth}
  \end{center}
  \caption{The ground-state mass of the magnetic gluelump for various
    actions, shown in units $\beta M$ against $1/\beta$.
    Note this is divergent, so one cannot trivially take a continuum
    limit.}
  \label{fig:gm}
\end{figure}

\begin{figure}
  \begin{center}
    \psfig{file=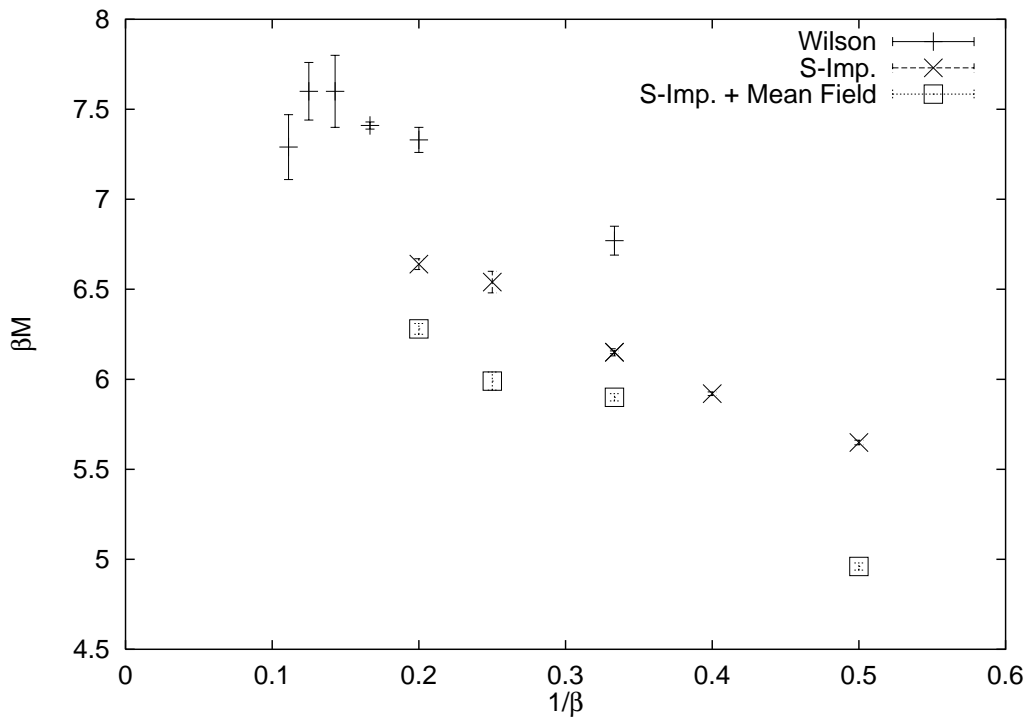,angle=270,width=\figwidth}
  \end{center}
  \caption{As \protect\ref{fig:gm}, but for the ground-state mass of the
    electric gluelump.}
  \label{fig:ge}
\end{figure}

For the gluelump masses, the results are shown in
table~\ref{tab:glmasses}, and the product of the masses of the
lowest-lying magnetic and electric gluelumps with $\beta$ are plotted
in figures~\ref{fig:gm} and~\ref{fig:ge} respectively; the comment
about $\xi$ above still applies.  As already noted, these are
divergent and one cannot take a simple continuum limit.  Given this,
the fact that the graph for the Wilson action is extremely flat ---
even in comparison with the string tension --- is the more surprising,
since it is highly unlikely this comes from a coincidental
cancellation between lattice artifacts and the divergent contribution.
So it would seem both effects are in this region benign.

However, we are not in a position to say the same thing about the
improved results, which are considerably shifted from the Wilson
values, and, though they show a rough plateau at small $\beta$, drift
upwards for large $\beta$, more noticeably in the case of the electric
gluelump.  The latter is also less well measured, due partly to the
higher mass, but also possibly to the operators being better optimised
for the magnetic gluelump.  The cause of the drift is hard to
determine, given all the possible effects.

Furthermore, taken together the plots also indicate the ratio between
the magnetic and electric gluelumps at a given $\beta$ is not
maintained by the S-improved action for $\beta\lesssim 3.0$, with or
without mean-field improvement.  All the graphs cast doubt on the
benefits of mean-field improvement at the lowest $\beta$, where the
results do not fit any obvious trend.

This complicates the choice of action, but in the absence of any
better suggestion we shall retain the S-improved action for results at
small $\beta$.

\section{Adjoint strings: mixing and breaking}

Given the results of the previous section, we decided to concentrate
on just one coupling in the Wilson case at $\beta=6.0$ and two with
the S-improved action, $\beta=3.0$, where scaling in the gluelump
sector appears to be just breaking down, and $\beta=2.0$, where we
certainly cannot trust scaling.  Furthermore, $\beta=2.0$ seemed too
small to provide additional mean-field improvement, at least in the
way implemented here, and we have kept $u_s=1$ for both couplings.

The runs are shown in table~\ref{tab:runs}.  We have used three sets of
operators: Wilson loops (\OpW), magnetic dumbbells (\OpM), and electric
dumbbells (\OpE), with the corresponding combinations for the off-diagonal
elements of the mixing matrix.  For each operator we used three levels of
fuzzing: one was always the raw (unfuzzed) operator (denoted by adding 0 to
the operator formed out of the unfuzzed links), one had a small amount of
fuzzing (denoted 1), and the third consisted of the optimal amount of
fuzzing (denoted 2).  The two sets of fuzzing parameters we have used in
each case are shown in the table; it should be noted that overlaps with the
gluelump states and with the unbroken string are not necessarily optimised
by the same set of fuzzing parameters, and our choice was to concentrate on
the unbroken string.  Hence the full matrix is $9\times9$, which we
abbreviate by indicating the operators and fuzzing levels included as
\OpW\OpM\OpE012; we shall use this as a shorthand when indicating the
basis. Likewise, \OpW012 indicates a basis of the three levels of fuzzing
but from Wilson loops alone; \OpW2 indicates the single most highly fuzzed
Wilson loop operator, and so on.

In two cases, once mixing had been clearly seen with the
full matrix, we performed further runs with only the Wilson loop operators
to try to see if these had some finite overlap with the broken string,
which is why we give two sets of configuration numbers.

For the $\beta=6$ run we simply used all the spatial separations along
the axis from $R=1$ to 16.  We used preliminary results from this to
concentrate points in the expected string-breaking region for the two
coarser lattices, where we have also used both on- and off-axis points
for the spatial separation $R$ in both the Wilson loop and
two-dumbbell operators.

\begin{table}
\begin{center}
\begin{tabular}{|rrllllll|}
\hline
& & & & \multicolumn{2}{c}{Fuzzing $(n_f, c_f)$} & & \\
$\beta$ & Action & $\xi$ & Size & Level 1 & Level 2
& Full configs. & Loop configs. \\
\hline
2.0 & S & 0.25 & $8^2\times32$  & $(1, 10.0)$ & $(4, 10.0)$ & 4500 & 8500 \\
3.0 & S & 0.25 & $12^2\times36$ & $(2, 5.0)$ & $(6, 5.0)$   & 2000 & 2000 \\
6.0 & W & 1    & $32^3$         & $(5, 2.3)$ & $(30, 2.3)$  & 2200 & 5900 \\
\hline
\end{tabular}
\null\vspace{12pt}
\begin{tabular}{|rl|}
\hline
Operators & Notation \\
\hline
Wilson loop & \OpW \\
`Dumbbell' with magnetic quantum numbers & \OpM \\
`Dumbbell' with electric quantum numbers & \OpE \\
\hline
\end{tabular}
\end{center}
\caption{Runs performed in the search for adjoint string breaking.
We denote the three types of operator included by the symbols shown, and
indicate the fuzzing level by adding $0$ (no fuzzing), $1$ (the first set
of parameters given above) or $2$ (the second set).  Hence our full set of
operators for each run is denoted \OpW\OpM\OpE012, which implies we have
used a $9\times9$ mixing matrix of the form shown schematically in
fig.~\ref{fig:mixing}.}
\label{tab:runs}
\end{table}

First we need to choose a diagonalisation basis $(T_D,T_D+1)$ as
described above.  $T_D=0$ is not a useful choice, as the Wilson loops
all have area zero and hence the matrix $M(R,0)$ consists of many
degenerate elements equal to unity, so we have tried $T_D=1$ and
larger.  In principle we could pick $T_D$ separately for each
diagonalisation.  In practice we have found no case where $T_D>1$
improves the result of the calculation, and many are made unstable by
this choice --- particularly those where the errors on $C(R,T_D)$ are
significant, which are just those cases we are most interested in
seeking an improvement.  Hence all our results use $T_D=1$.

Our effective mass plateaus thus obtained are in most cases very good,
with $m_\mathrm{eff}(t=2)$ already close to the plateau value for the
Wilson data, and we are clearly able to calculate the two lowest
masses in the majority of cases.  In the results shown, we have
decided point by point at what time value the fit to the exponential
should start by considering both the $\chi^2$ per degree of freedom
and the flatness of the plateau, to try to be reasonably but not
excessively conservative.  All use either $t=2$, 3 or 4.

\subsection{Results from full mixing matrix}

\begin{figure}
  \begin{center}
    \psfig{file=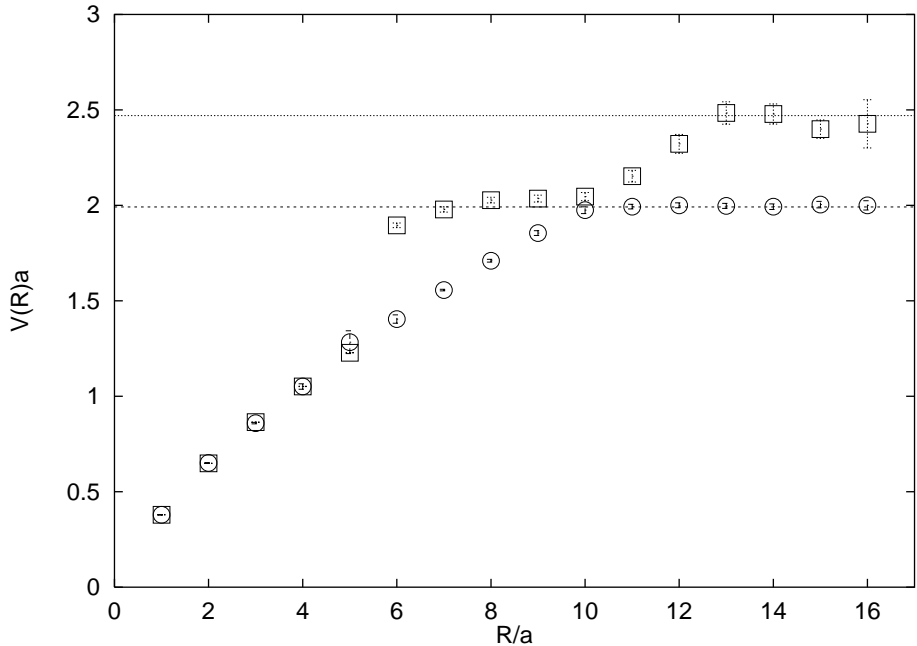,angle=270,width=\figwidthless}
  \end{center}
  \caption{The results for the adjoint potential including all operators
    (\OpW\OpM\OpE012) at $\beta=6.0$, Wilson action, showing the first two
    states, as well as the twice the masses of the two lowest gluelumps.
    Diagonalisation has failed for the excited state at small distances.}
  \label{fig:w6_full}
\end{figure}

\begin{figure}
  \begin{center}
    \psfig{file=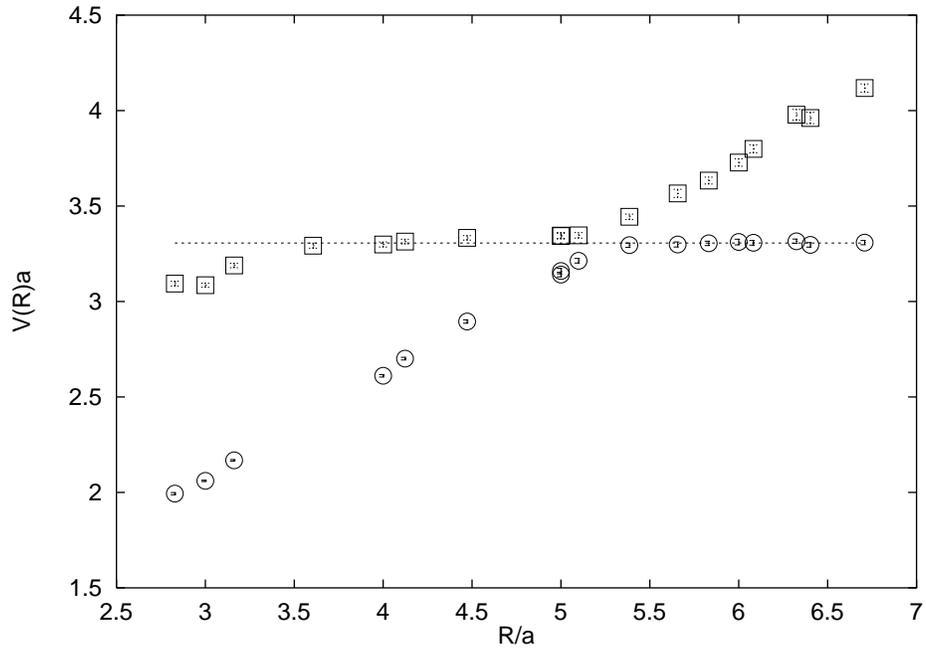,angle=270,width=\figwidthless}
  \end{center}
  \caption{The results for the adjoint potential including all
    operators (\OpW\OpM\OpE012) at $\beta=3.0$, improved action showing the
    first two states, as well as the twice the mass of the lowest magnetic
    gluelump.  Note the narrower range of distances in comparison with
    fig.~\protect\ref{fig:w6_full}.}
  \label{fig:sp3p_full}
\end{figure}

\begin{figure}
  \begin{center}
    \psfig{file=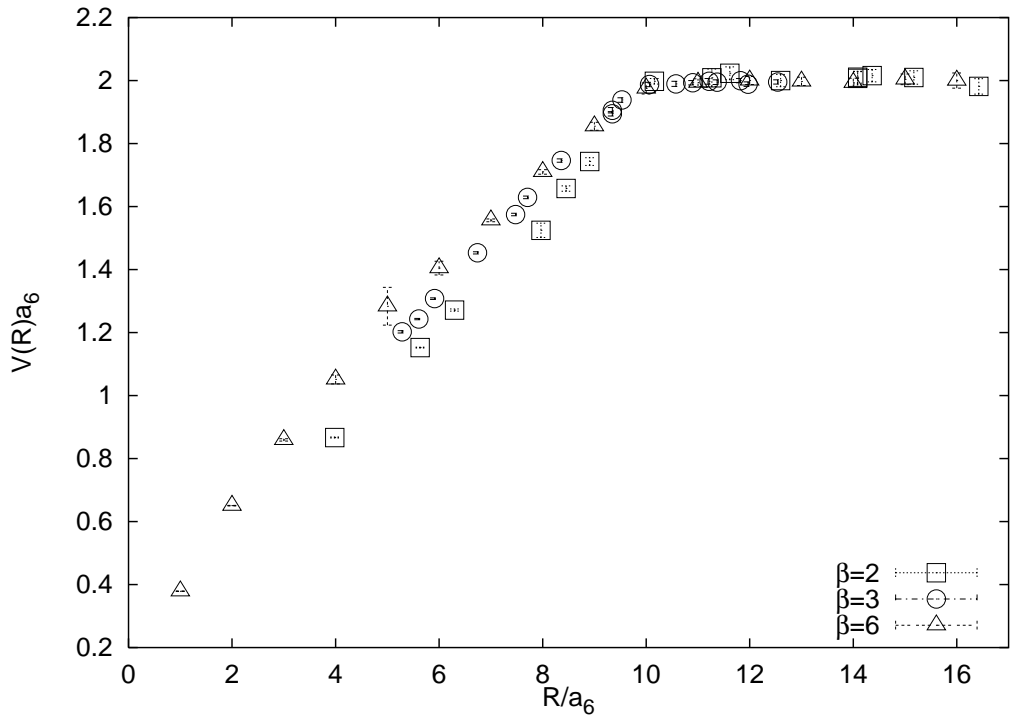,angle=270,width=\figwidth}
  \end{center}
  \caption{The lowest state adjoint potential from all operators
    (\OpW\OpM\OpE012), showing all three sets of data rescaled to $\beta=6$
    as described in the text.}
  \label{fig:full_com}
\end{figure}

The results for the two lowest-lying states of the adjoint potential
from the complete basis of 9 operators (\OpW\OpM\OpE012) are shown in
fig.~\ref{fig:w6_full} (Wilson at $\beta=6.0$) and
fig.~\ref{fig:sp3p_full} (S-improved at $\beta=3.0$); in each case,
twice the mass of the lowest-lying gluelump or gluelumps is shown as a
straight line.

The complexities of scaling referred to above make it hard to overlay
these graphs.  First we rescale all points according to the string
tension; this leaves the matter of the divergence of the two static
sources, corresponding to a vertical shift of the potentials at
different $\beta$.  As the plateau can be clearly seen to correspond
to the mass of two magnetic gluelumps, we have then shifted the
vertical axis by eye until the plateaus roughly correspond (we do not
here need the extra accuracy a full fit would give).  With good
scaling behaviour, this procedure should be qualitatively correct.
This is attempted in fig.~\ref{fig:full_com}; the labels of the axes
show the values for the $\beta=6$ run.  The point at which the string
breaks seems to be roughly universal, although the rising parts of the
potential do not agree very well.

At $\beta=6.0$ the diagonalisation has failed for $R\lesssim5$, where
it was not able to separate the first excited state from the ground
state; a similar tail appears in the $\beta=3.0$ data, which again
probably represents failure of the procedure rather than anything
physical (note fig.~\ref{fig:sp3p_full} does not go to such short
distances as fig.~\ref{fig:w6_full}).  A possible clue as to why this
should happen is that the `U'-shaped operators forming the
off-diagonals in fig.~\ref{fig:mixing} have themselves a strong
overlap with the unbroken string at short distances, approaching
unity as $R\to0$, hence it may become more difficult to resolve the
excited two-gluelump state from the full set of operators; however,
even the two-dumbbell operators alone have difficulties in this
region, as described below.

Notwithstanding the problems, the overall effect is clear; one sees
not one but two string breakings, as the first excited potential
itself flattens off at the value of twice the mass of the lowest-lying
electric gluelump.  The effect is very stark; there seems to be very
little sign of `repulsion' between the two curves at the crossover
point, as is commonly associated with such phenomena, a point already
noted in ref.~\cite{Mi92}, although the picture is qualitatively
similar to that seen in breaking of a fundamental string to two scalar
mesons~\cite{PhWi98,KnSo98}.  It is clear here, as there, that the mass of
the particle in the broken state is the major factor determining the
breaking scale, which can be read of in lattice units from the
horizontal axis.

We performed initial fits during the diagonalisation; these give us a
way of estimating the overlap of the original operators with both of
the first two states.  Since two-exponential fits are notoriously
difficult, we first fit a single exponential to each of the lowest two
eigenmodes; these are essentially the same fits as those used to
extract the potential itself.  We then make a two-exponential fit to
the original data sets for the individual operators, but with the
masses fixed to the values obtained from the first fit so that only
the two coefficients are free.  This will not work if one has reason
to think the data in question have a significant overlap with a third
exponential; we have not actually identified any third state to see if
this is the case.  Of course, the whole diagonalisation procedure is
anyway at the mercy of effects from badly resolved states.

\begin{figure}
  \begin{center}
    \psfig{file=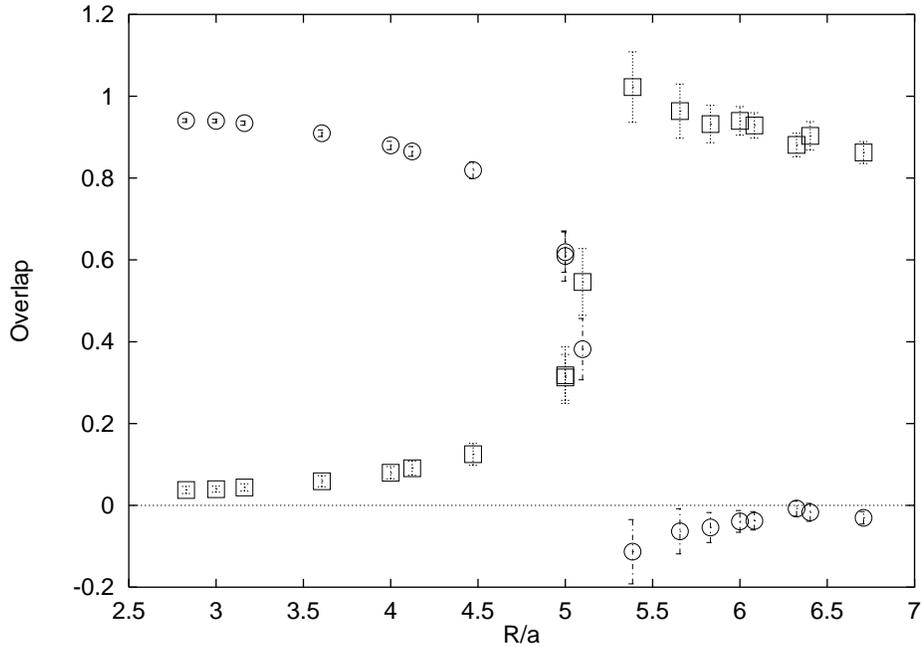,angle=270,width=\figwidthless}
  \end{center}
  \caption{Overlap of the most highly fuzzed Wilson-loop operator (\OpW2)
    with the ground state (circles) and first excited state (squares)
    of the adjoint potential for $\beta=3.0$.}
  \label{fig:proj_loops}
\end{figure}

\begin{figure}
  \begin{center}
    \psfig{file=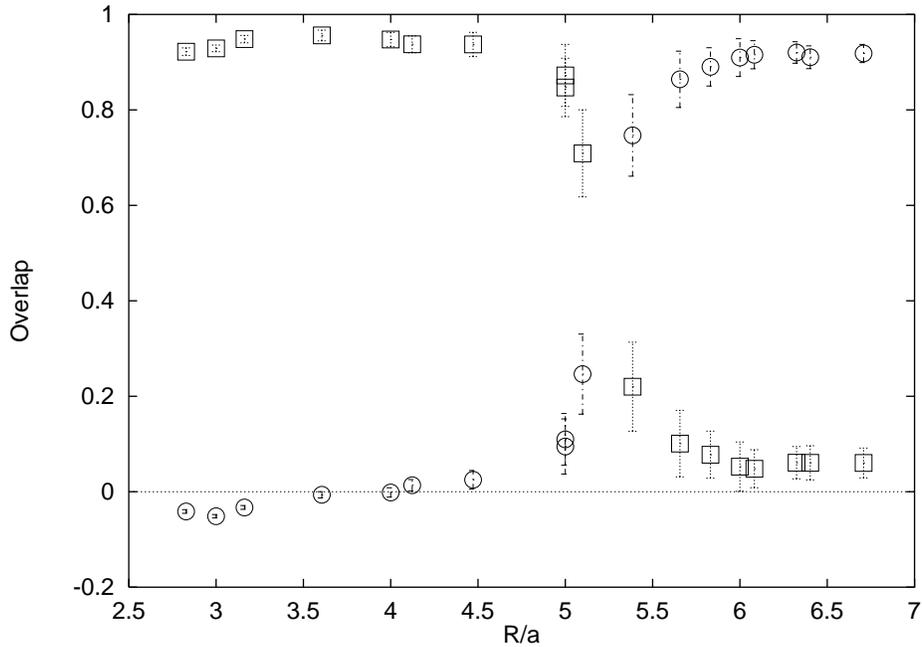,angle=270,width=\figwidthless}
  \end{center}
  \caption{As fig.~\ref{fig:proj_loops}, but showing the overlaps of
    the most highly fuzzed two-magnetic-dumbbell operator (\OpM2).}
  \label{fig:proj_mag}
\end{figure}

This has worked most successfully for the $\beta=3.0$ data, possibly
because the combination of coupling and asymmetry give it the longest
plateaus.  In fig.~\ref{fig:proj_loops} we show the overlap of the
most highly fuzzed loop operator (\OpW2) with the ground state and the first
excited state of the potential, and in fig.~\ref{fig:proj_mag} we show
the corresponding graph for the operator consisting of two magnetic
dumbbells, again with the highest fuzzing (\OpM2).  As one would expect, these
complement one another nicely, with a clear crossover around $R=5.2a$.
The crossover of the overlaps appears less abrupt than that in the
states themselves, again in accordance with the results of
ref.~\cite{PhWi98}.  In the region where the states have similar
masses, it may be difficult for the diagonalisation to separate the
first two states; this presumably explains why in
fig.~\ref{fig:proj_loops} immediately after string breaking the
overlap for the excited state is too high and that for the ground
state too low.

\subsection{Results from individual types of operator}

The problem that started this project was the difficulty (even
apparent impossibility) of seeing string breaking in Wilson loop
operators alone and figs.~\ref{fig:proj_loops} and~\ref{fig:proj_mag}
have confirmed our expectations that each operator is linked
particularly with either the unbroken or the broken string, but not
both.  It is therefore interesting to see if one can detect any finite
overlap, however small, of the operators with the `wrong' states.

\begin{figure}
  \begin{center}
    \psfig{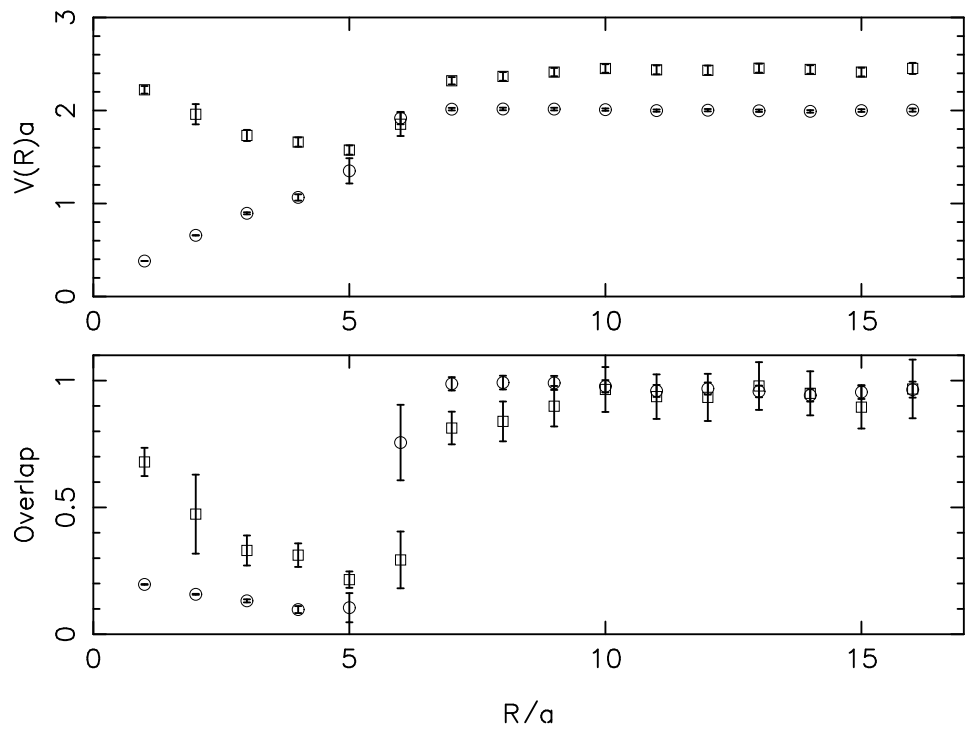}
  \end{center}
  \caption{Above, the adjoint potential from diagonalisation of the three
    operators consisting of two magnetic dumbbells (\OpM012) at $\beta=6$ for
    the ground state (circles) and first excited state (squares).
    Below, the overlap with the states resolved.}
  \label{fig:w6_mproj2}
\end{figure}

It turns out that the overlap of the two-dumbbell operator with the
unbroken string at small distances is large enough to produce quite
reasonable results.  The behaviour for the three operators consisting
of two magnetic dumbbell at $\beta=6.0$ (\OpM012) is shown in
fig.~\ref{fig:w6_mproj2}; here we have diagonalised just the three
relevant data channels in the mixing matrix --- note the difference
from fig.~\ref{fig:proj_mag} where we showed the overlap for a
single operator of this type.  For the lowest state, the change
between projecting the unbroken and broken string is very clear, with
large errors in the region where it is hard to separate the two
states; in particular, the unbroken string is not followed for
$R\gtrsim5a$, as it should intersect the two-gluelump plateau around
$R=10a$ (compare fig.~\ref{fig:w6_full}).

For the overlaps with the state resolved, also shown in
fig.~\ref{fig:w6_mproj2}, this time we have simply used the
coefficient of a single exponential fit.  At small $R$, where the
string-like state is resolved, the overlap starts at around $20\%$
and decreases, then increases suddenly where the broken state
dominates.  This is, indeed, our most striking evidence that
particular operators can have significant overlaps with both the
broken and unbroken string.  A similar picture appears for the other
data sets, but is less striking as the points are concentrated around
the region of string breaking.

We have already seen that with the complete basis the procedure does not
work well for the excited state at short distances, so it is
questionable whether the corresponding part of fig.~\ref{fig:w6_mproj2}
shows a physical state, rather than merely an artifact of the method:
the overlap is much lower than that where the two-gluelump state is well
resolved, although higher than with the ground state.  The
diagonalisation procedure is unable to resolve the two states
separately in the crossover region where the string-like and broken
states have similar masses.  For large $R$, where the operators resolve
two broken-string states, the mass of the excited state corresponds to
twice the lowest electric gluelump mass, two magnetic and two electric
gluelumps having the same quantum numbers.

\begin{figure}
  \begin{center}
    \psfig{file=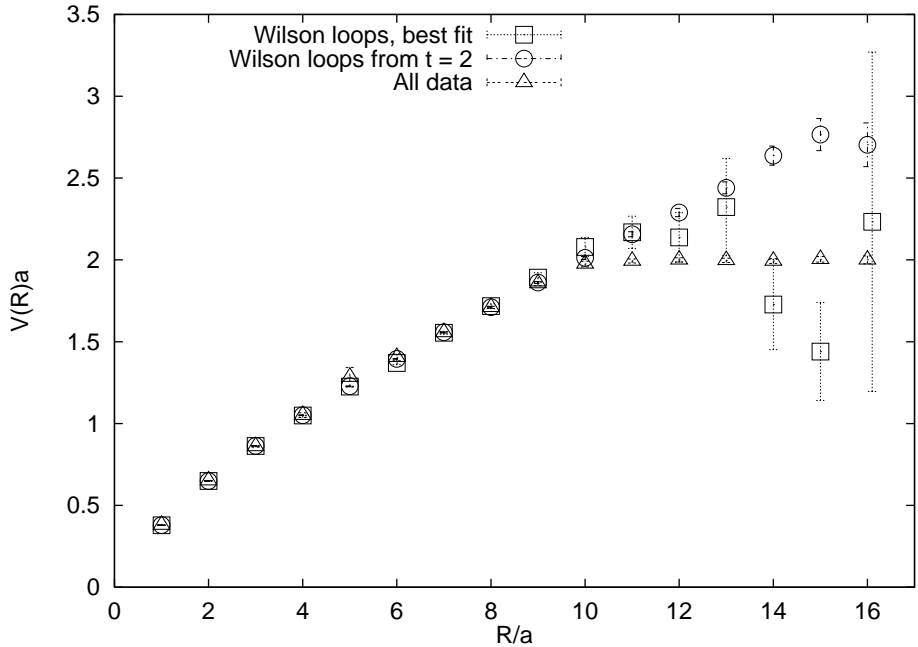,angle=270,width=\figwidthless}
  \end{center}
  \caption{The Wilson loop sector, operators \OpW012, compared with
    the full results, operators \OpW\OpM\OpE012, for $\beta=6$.}
  \label{fig:w6_l_break}
\end{figure}

A finite overlap of the Wilson loops with the broken state, on the
other hand, is much harder to see.  We have only succeeded in one
case, for $\beta=6.0$ --- so our attempts to use coarser lattices to
improve overlaps appear to have been in vain, in contrast to the
case of the fundamental representation with dynamical fermions in
ref.~\cite{Tr98}.  Even here, the evidence would be distinctly
ambiguous if we did not have the full operator basis for comparison.
In fig.~\ref{fig:w6_l_break} we show the fits from the results
diagonalised using Wilson loops alone, i.e. the basis of three operators
\OpW012, for the optimum plateau region,
usually starting at timeslice 3 (the last point has been slightly
shifted to the right for clarity); then the same results but fit from
timeslice 2; and finally the full results with all operators
(\OpW\OpM\OpE012).  With all this information, it becomes clearer that a
small overlap to the broken state eventually overcomes a larger overlap to
the unbroken state, yielding a potential with large errors --- the size of
the errors alone alerts one to the fact that a second state is present, as
the errors in the case where only the unbroken string is seen are small.

\begin{figure}
  \begin{center}
    \psfig{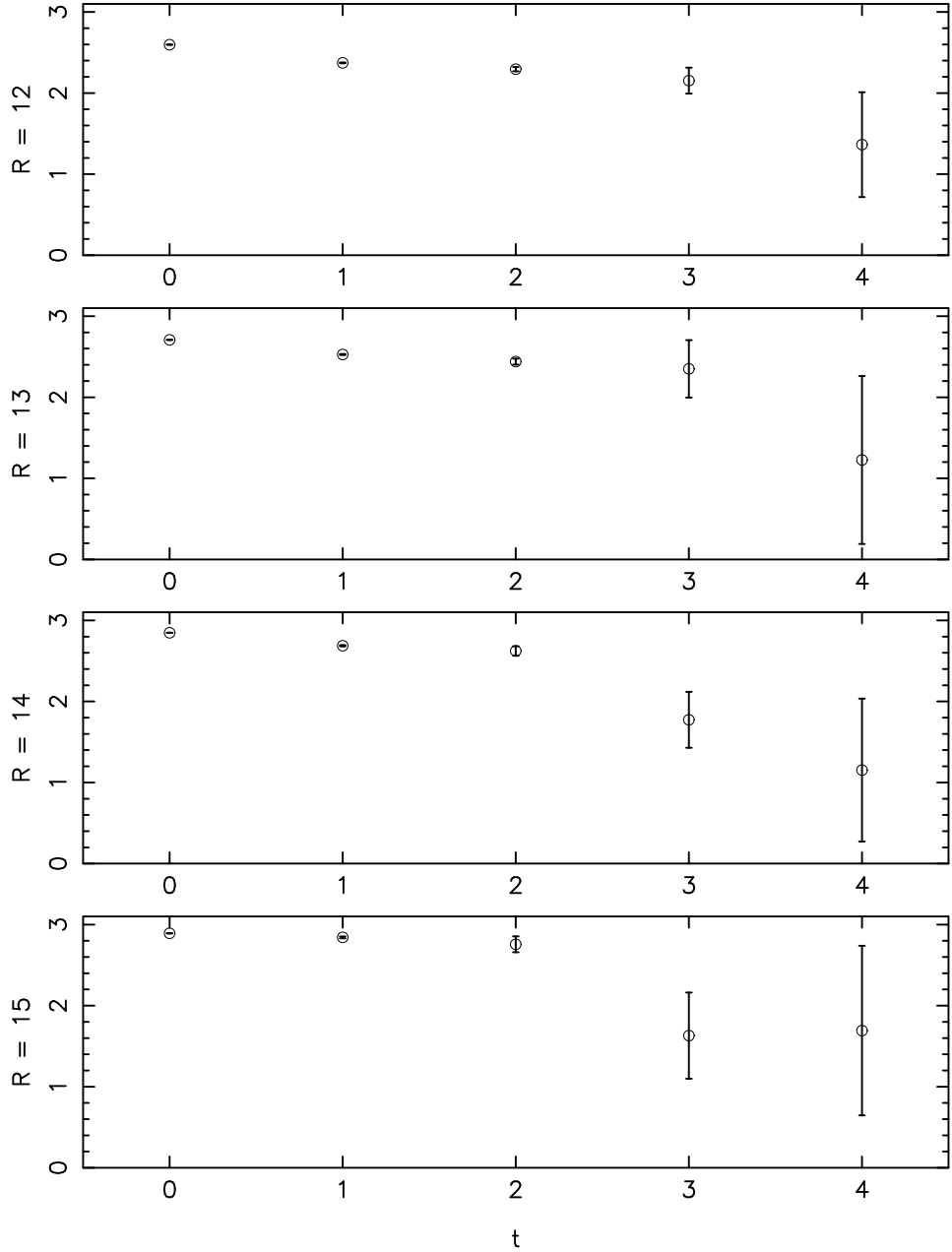}
  \end{center}
  \caption{The effective mass plots for four of the adjoint potentials
    from Wilson loops (operators \OpW012) shown in
    \protect\ref{fig:w6_l_break} which lie just after the region of string
    breaking.}
  \label{fig:w6_em}
\end{figure}

To see what is happening in more detail, we show a sequence of
effective mass plots for the region where string breaking occurs in
fig.~\ref{fig:w6_em}.  As $R$ increases, one sees the plateau begin to
drop for $t=3$ and the error bar increase substantially.  The
coefficient of the exponential fit in this region is too poorly
determined for an estimate of the overlap, but it is certainly no more
than about five to ten percent.  A similar effect is already visible
using the most highly fuzzed Wilson loop operator on its own.  These
results accord with the suggestion that the Wilson loop has a finite
but poor overlap to the broken state~\cite{KnSo98}.

\begin{figure}
  \begin{center}
    \psfig{file=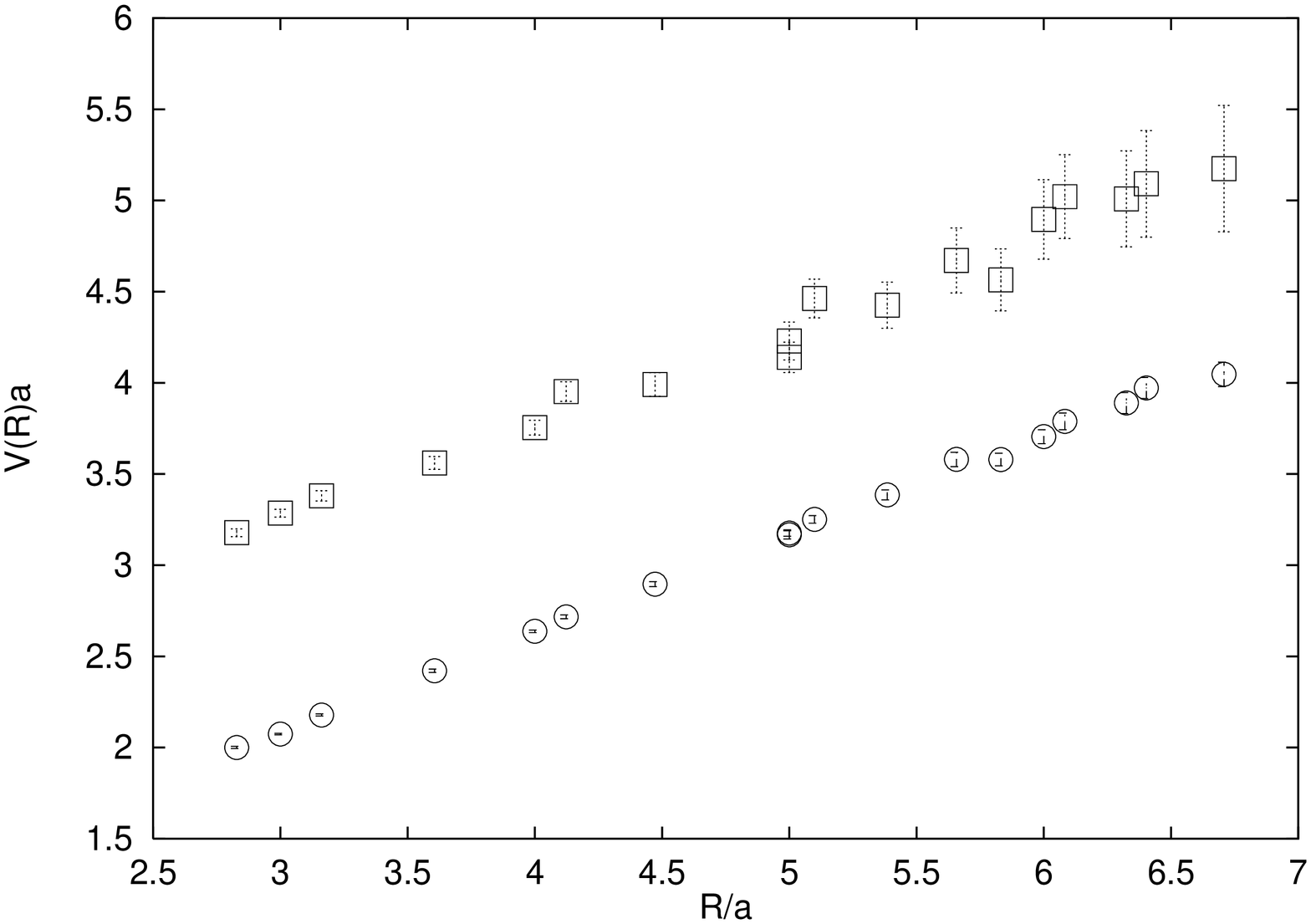,angle=270,width=\figwidth}
  \end{center}
  \caption{The first two states of the adjoint potential from the
    three Wilson loop operators (\OpW012) alone at $\beta=3.0$.  String
    breaking, if visible, would begin at around $R=5.2a$.}
  \label{fig:sp3p_loops}
\end{figure}

For comparison, we show also the first two states of the adjoint potential
at $\beta=3.0$ using the Wilson loops alone (operators \OpW012) in
fig.~\ref{fig:sp3p_loops}; here and at $\beta=2.0$ there is no sign of
string breaking, nor any indication from the errors of that another state
may be present, nor are there any obvious anomalies in the effective
masses; this is just the sort of picture usually obtained up to now which
has caused unquiet~\cite{DiPe98}.  The reason why there is no sign of
string breaking in this case is not clear, and is probably hard to
determine given we are dealing with very low overlaps in any case.  One
possibility is that the fuzzing is not so well optimised on the coarser
lattices.

\subsection{Comparison with the fundamental potential}

\begin{figure}
  \begin{center}
    \psfig{file=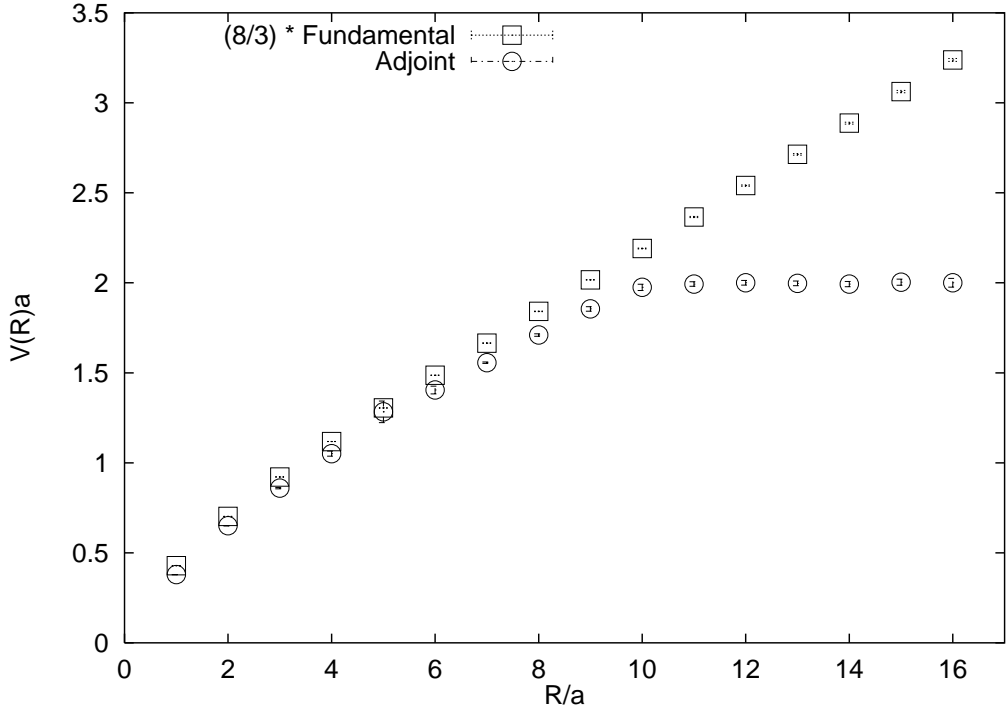,angle=270,width=\figwidth}
  \end{center}
  \caption{The adjoint potential from all operators shown in
    comparison with the fundamental potential rescaled by a factor
    $8/3$ at $\beta=6$.}
  \label{fig:w6_fund_adj}
\end{figure}

We also show the fundamental potential for comparison; because of the
complexities of scaling, we shall just give results just for $\beta=6$.  In
fig.~\ref{fig:w6_fund_adj}, we show the adjoint potential from all
operators (\OpW\OpM\OpE012) compared with $8/3$ times the fundamental
potential (calculated with the operators \OpW012 but with the fundamental
rather than the adjoint trace), derived from the ratio of quadratic Casimir
operators; the so called `Casimir scaling' obeying this rule has long been
known to hold very closely in 3+1 dimensions.  As found
previously~\cite{PoTr95}, in 2+1 dimensions the fundamental line lies
slightly higher.  The natural interpretation is that there is some form of
precocious screening of the adjoint potential by virtual gluons.  Here one
sees an apparently increasing divergence as the string breaking region is
approached which lends some credence to this point of view.

\section{Summary}

We have looked at the breaking of the adjoint string in pure SU(2)
gauge theory in 2+1 dimensions where the potential is expected to
saturate into two `gluelumps', states consisting of both static and
dynamical gluons.  We have used a basis of operators which includes
both Wilson loops, corresponding to creation and annihilation of an
unbroken string, and `two-dumbbell' operators, corresponding to the
same process for two localised gluelumps a certain distance apart.
Diagonalising the whole basis of operators shows a clear and
unambiguous string breaking.  This is not surprising given that one
knows that there will be a two gluelump state for large enough
separations of the sources, and that eventually the increasing
potential will overtake it so that this will be the ground state.  The
onset of this breaking is rather sudden; the crossover of the states
seems to be abrupt.

We have also investigated the overlap of the different types of
operator with the different states.  Using fuzzing the overlap of the
Wilson loop with the string state and of the dumbbell operators with the
gluelump state is excellent, well over 90\%.  The two-dumbbell
operator has at short distances a significant overlap, up to around
20\%, with the unbroken string.  This is a pleasant surprise, as it
indicates that the string-like potential is not simply an artifact of
Wilson loop operators~\cite{DiPe98}.  A physical picture is that when
the fields associated with the two dumbbells are close together their
fields can overlap and hence are able to overcome the screening.  We
have found one case where a Wilson loop basis has some overlap with
the broken-string state, but it is so weak that we cannot estimate it.
Large distances, where the unbroken string has a very high energy and
hence disappears at shorter times, and large statistics would be
required to pin this down.

Scaling between different couplings, and even more between different
actions, is problematic because of divergences in the gluelump, and
because we have used very coarse lattices to see if we can improve the
overlap.  These coarser lattices do not seem any more revealing; nor
do we expect that increasing $\beta$ would reveal new physics as
scaling for quantities with a proper continuum limit appears unproblematic
for $\beta>6$~\cite{Te98}.

In looking for Casimir scaling between the fundamental and adjoint
potentials, we appear to see an increased screening of the
adjoint potential as the region of string breaking is reached.

These results, taken together with the results for breaking of the
string between fundamental sources in ref.~\cite{Tr98}, support the
view~\cite{Gu98} that the main issue is to find combinations of
operators with high overlaps to the physical states present, and are
evidence against (though do not necessarily rule out) the more alarming
suggestions that the potential seen in lattice simulations might be
unphysical~\cite{DiPe98}.

We would expect the features discussed to be qualitatively similar with
other SU($N$) gauge groups and in 3+1 dimensions; our results suggest
that the naive estimate for the point of string breaking, where the mass
of the two-gluelump state intersects the rising potential, will be a
good one.

Almost simultaneously with this paper, another group reported very
similar conclusions about string breaking in the same
model~\cite{PhWi99}.

\section*{Acknowledgments}

I would like to thank Colin Morningstar and Hartmut Wittig for
pointing out the difficulties with renormalisation of the space-time
asymmetry, and of the gluelump masses, respectively.

\end{document}